\documentclass[runningheads]{llncs}
\usepackage{graphicx}
\usepackage{todonotes}

\newcommand\blfootnote[1]{%
  \begingroup
  \renewcommand\thefootnote{}\footnote{#1}%
  \addtocounter{footnote}{-1}%
  \endgroup
}

\begin{document}
\title{Topology-preserving augmentation for CNN-based segmentation of congenital heart \\ defects from 3D paediatric CMR
\thanks{Nick Byrne is funded by a National Institute for Health Research (NIHR), Doctoral Research Fellowship for this research project. This report presents independent research funded by the NIHR. The views expressed are those of the author(s) and not necessarily those of the NHS, the NIHR or the Department of Health and Social Care. The authors have no conflicts of interest to disclose.}}

\titlerunning{Topology-preserving augmentation for 3D segmentation}
%

\author{Nick Byrne\inst{1,2}
\and
James R. Clough\inst{2}
\and
Isra Valverde\inst{2,3}
\and
\\ Giovanni Montana\inst{4}$^{\dagger}$
\and
Andrew P. King\inst{2}$^{\dagger}$
}

\authorrunning{N. Byrne et al.}

%

\institute{Medical Physics, Guy's and St. Thomas' NHS Foundation Trust, London, UK \and
School of Biomedical Engineering \& Imaging Sciences, King's College London, UK \and
Paediatric Cardiology, Guy's and St. Thomas' NHS Foundation Trust, London, UK \and
Warwick Manufacturing Group, University of Warwick, Coventry, UK
\email{nicholas.byrne@kcl.ac.uk}}

\maketitle              

\begin{abstract}

Patient-specific 3D printing of congenital heart anatomy demands an accurate segmentation of the thin tissue interfaces which characterise these diagnoses.
Even when a label set has a high spatial overlap with the ground truth, inaccurate delineation of these interfaces can result in topological errors.
These compromise the clinical utility of such models due to the anomalous appearance of defects.
CNNs have achieved state-of-the-art performance in segmentation tasks.
Whilst data augmentation has often played an important role, we show that conventional image resampling schemes used therein can introduce topological changes in the ground truth labelling of augmented samples.
We present a novel pipeline to correct for these changes, using a fast-marching algorithm to enforce the topology of the ground truth labels within their augmented representations.
In so doing, we invoke the idea of cardiac contiguous topology to describe an arbitrary combination of congenital heart defects and develop an associated, clinically meaningful metric to measure the topological correctness of segmentations.
In a series of five-fold cross-validations, we demonstrate the performance gain produced by this pipeline and the relevance of topological considerations to the segmentation of congenital heart defects.
We speculate as to the applicability of this approach to any segmentation task involving morphologically complex targets.
\blfootnote{\hspace{-2.5mm}$^{\dagger}$ \emph{Joint last authors.}}

\keywords{Image segmentation  \and Data augmentation \and Topology.}
\end{abstract}

\section{Introduction}

\setcounter{footnote}{0}

Medical image segmentation is an integral part of many pipelines for the analysis of clinical data. For many applications, such as the calculation of ventricular volumes, algorithmic approaches need only achieve a segmentation that shares a sufficient overlap with an expert defined reference standard.
This can be assessed using the Dice Similarity Coefficient (DSC).
However, in other cases the topology of the segmentation is also important.
For example, topologically correct segmentation is a prerequisite for the detailed visualisation of paediatric congenital heart disease (CHD) anatomy using patient-specific 3D printed models.

\begin{figure}[t]
\begin{center}
\includegraphics[width=\textwidth]{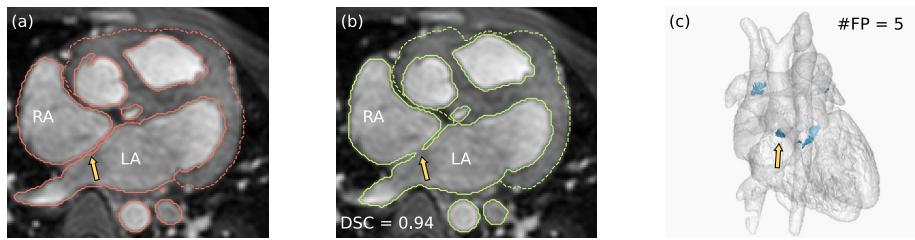}
\end{center}
\caption{An error in the segmentation of thin tissue interfaces such as the septum between left and right atria (LA, RA) can give rise to topological changes and the anomalous appearance of a congenital heart defect (yellow arrow). Whilst the DSC between ground truth (orange) and inferred (green) blood pool is high, (c) demonstrates the presence of five topologically and clinically relevant segmentation errors.} \label{interfaces}
\end{figure}

Segmentation of the congenitally malformed heart from CMR images is a challenging task due to
inhomogeneity in signal intensity, limited contrast-to-noise ratio and the presence of image artefacts \cite{pace15}.
Furthermore, significant variation in the structural presentation of disease limits the success of conventional methods such as atlas-based strategies \cite{zuluaga17}.
Finally, patient-specific 3D printing demands a high fidelity representation of disease, demonstrating anatomy at the limit of spatial resolution.
Segmentation results should accurately represent clinically meaningful thin tissue interfaces such as the atrial septum (see figure \ref{interfaces}).
Inaccurate interface segmentation introduces anomalous topological features that may falsely indicate the presence of congenital heart defects.
Consequently, exponents of patient-specific 3D printed heart models have hitherto relied on manual and semi-automated segmentation methods, typically requiring at least an hour of manual interaction per patient \cite{pace15}.

Convolutional neural networks (CNNs) have been successfully applied to a multitude of image segmentation tasks, including the delineation of congenital heart defects from CMR data.
Wolterink et al. \cite{wolterink17} trained a slice-wise, 2D CNN using dilated convolutions.
Meanwhile, Yu et al. \cite{yu17} explored deep supervision \cite{yu17-frAQB} and dense connectivity within 3D CNNs.
Considering a limited training set of just ten cases, these approaches achieved impressive results in terms of spatial overlap.
However, automated approaches cannot yet match the overlap performance of the leading semi-automated procedures \cite{losel17,pace15}, and have largely paid little attention to topological correctness.

\begin{figure}[t]
\begin{center}
\includegraphics[width=\textwidth]{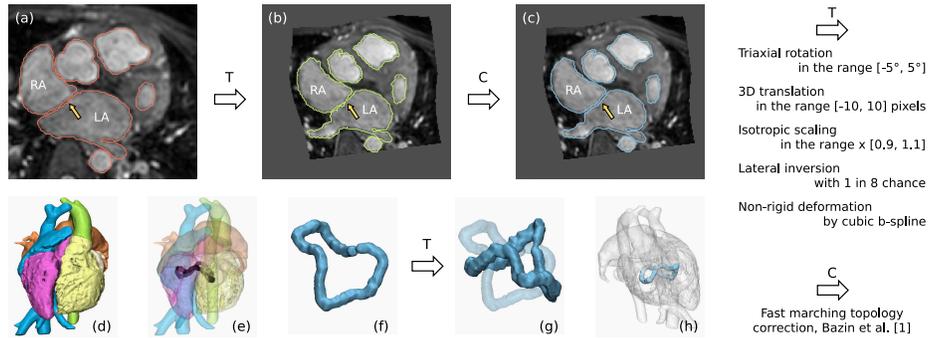}
\end{center}
\caption{Spatial transformation and nearest neighbour resampling of a ground truth label set (a), can result in anomalous topological changes such as defects within the atrial septum (b). Such changes can be corrected by consideration of a CCT template (f). Having two shunts between the respectively topologically spherical left and right heart (ventricular and atrial septal defect), this patient exhibits toroidal CCT. The template is derived by topological erosion of a multi-class representation of the blood pool (d, e) and subsequently transformed to the space of the augmented image (g, h).} \label{pipeline}
\end{figure}

Especially in the paediatric setting, developers of medical image segmentation algorithms cannot generally assume a database of thousands or millions of training cases.
Instead, state-of-the-art CNNs have relied on data augmentation schemes.
Augmentation acts as a source of regularisation and generalisation, capturing modes of variation likely to exist in the underlying population but which are absent from the training data.
Spatial scaling, small angle rotation and non-rigid deformation are attractive transformations for augmentation, accounting for variation in patient size, orientation and posture.
However, under such schemes, and when subsequently resampled by nearest neighbour interpolation, each of these can cause violations of ground truth topology near thin tissue interfaces (see figure \ref{pipeline}(b)).

Knowingly or otherwise, the best-performing previous work \cite{wolterink17,yu17,yu17-frAQB} has limited spatial augmentation to a subset of transformations that are topology-preserving: orthogonal rotation and lateral inversion only.
However, given that orthogonal rotation has no clinical rationale, this is unlikely to aid the generalisation of CNNs, providing a source of regularisation alone.

We hypothesise that topology-preserving label map augmentation is a pre-requisite to any advanced provision for topologically-informed deep learning.
To investigate this hypothesis we make the following contributions in the context of CHD segmentation from 3D CMR:

\begin{itemize}
    \item We present a novel pipeline for the augmentation of label map data in a topology-preserving manner.
    \item We present a novel metric for assessing the topological correctness of segmentation results, using it to demonstrate improved performance compared with previous work and with conventional image resampling schemes.
\end{itemize}

\section{Methods}

\subsection{Topology-preserving augmentation pipeline}

The notion of \emph{simple points}\footnote{Those whose binary label value can be flipped without changing the topology of the overall label map.} is central to the fast-marching topology correction tool developed by Bazin et al. \cite{bazin07}.
Starting from a scalar representation of the naively transformed object, this algorithm removes non-simple points from all isosurfaces, correcting the topology to match a known template.
Whilst correcting object topology to match that of a ball is straightforward, defining the template for morphologically complex congenital anatomy presents a greater algorithmic challenge.

Our solution (see figure \ref{pipeline}) is predicated on the idea that the whole heart blood pool has an uncomplicated topological representation.
In reality, the topology of the blood pool label can be highly complex, demonstrating numerous fine-scale features associated with the trabeculated, endocardial surfaces.
To avoid this complexity, we invoke a property that we refer to as cardiac contiguous topology (CCT).
This describes the structural relationships between sub-classes of the cardiac blood pool and their communication.
Importantly, the CCT captures the appearance of thin tissue interfaces and defects by defining how and where the heart's chambers and vessels are contiguous.

To remove topological features associated with trabeculation, each sub-class is corrected (via the  approach in \cite{bazin07}) to have topology equivalent to a ball at the outset. 
Once recombined, the topology of the blood pool is defined only by the connections of each cardiac sub-class.
Furthermore, we require that the blood pool class constitutes a well-composed set\footnote{The topology of a well-composed set is independent of neighbourhood connectivity.}.
Such label maps have the advantageous property that repeated topological erosion is guaranteed to result in a one voxel wide CCT template.
This captures the topology of the ground truth blood pool in a morphologically simple object free from thin interfaces (see figure \ref{pipeline}(f)).

Having established a CCT template for each ground truth label map, the two can be spatially transformed in tandem.
Whilst nearest neighbour resampling is likely to cause topological errors in the label map, the morphologically simpler CCT template can be resampled without incurring such changes.

We resample the transformed blood pool label using trilinear interpolation, realising an image bounded in the range $[0,1]$.
Akin to a probability map, this is corrected to share the topology of the transformed CCT template, ensuring that the arbitrary CCT of the ground truth labels is maintained (see figure \ref{pipeline}(c)).

Most often, the sub-valvular apparatus and its association with the papillary muscles are only partially visualised in CMR: the true-to-life topological properties of the myocardium are rarely apparent.
Hence, we resample the spatially transformed myocardium label by naive nearest neighbour interpolation.

\subsection{Study design}

\hspace{5mm}\textbf{Data} --- We employ the ten cases provided during the training phase of the HVSMR Challenge 2016 (see \cite{pace15} for acquisition and clinical details).
Each case includes an isotropic, high-resolution, axially-reformatted, 3D CMR volume acquired at Boston Children's Hospital and demonstrating CHD anatomy.
This is tightly cropped around a provided set of manually segmented labels, delineating the whole heart blood pool and myocardium as two separate classes.

Prior to experimentation, an expert in paediatric, CHD segmentation corrected small topological errors that were present in the provided blood pool label maps.
The vast majority of corrections removed false positive voxels from within thin interfaces.
Totally, 0.098\% and 0.0059\% of the blood pool and myocardium classes were changed respectively.
In a series of five-fold cross-validations (train on eight, test on two), we address a three-class segmentation problem, separating the blood pool, myocardium and background classes.
In the context of a deep CNN, the performance of the topology-preserving augmentation pipeline is compared with a naive, nearest neighbour resampling of label data and with augmentation by orthogonal rotation and lateral flipping only (as in \cite{wolterink17,yu17,yu17-frAQB}).

\begin{figure}[t]
\begin{center}
\includegraphics[width=0.8\textwidth]{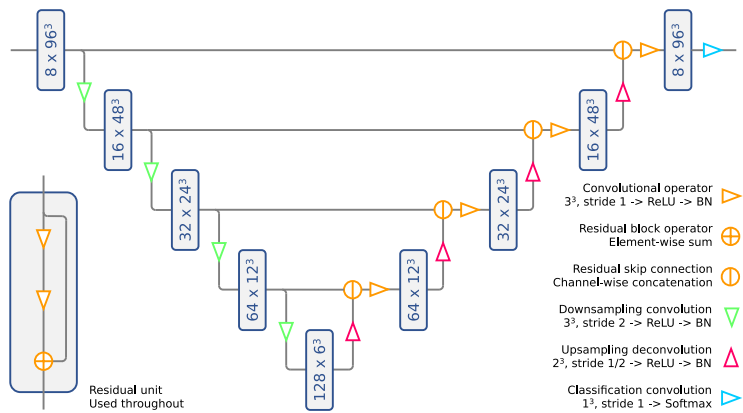}
\end{center}
\caption{The V-net architecture used. Output feature map sizes at training are indicated.} \label{architecture}
\end{figure}

\textbf{Architecture} --- We adopt the V-net architecture (see figure \ref{architecture}) in all experiments \cite{milletari16}, using 3D convolution to learn residual features across spatial scales.

\textbf{Metrics} --- Overlap performance was assessed using the DSC. However, such global metrics are largely insensitive to errors in thin interface regions.
To characterise the topology of the inferred blood pool, we introduce a novel, interpretable metric.
This extends the notion of simple points to connected components, algorithmically counting the number of topologically relevant clusters of voxels where inferred and ground truth segmentations disagree (see figure \ref{interfaces}(c)).
This approach is clinically meaningful as clusters of topologically relevant errors indicate the anomalous
appearance of congenital defects.

\textbf{Implementation} --- Prior to augmentation, all CMR data were normalised to have zero mean and unit variance.
The topology-preserving augmentation pipeline used the SimpleITK package for spatial transformation and resampling; fast-marching topology correction \cite{bazin07} and associated topological operations used relevant plugins for the Medical Image Processing And Visualisation (MIPAV) platform.
From the ten cases provided by the HVSMR Challenge, a total of 10,000 training examples were pre-computed by data augmentation according to figure \ref{pipeline}.
For comparison we pre-computed a further 10,000 training examples by augmentation using orthogonal rotation and lateral inversion alone. In both cases we also made small perturbations to the voxel intensity of the image data.

All models were trained for 8,000 iterations using the Adam optimiser (Pytorch default settings for learning rate and beta parameters) and the categorical cross entropy loss.
Each batch contained eight image patches of $96\times96\times96$ voxels, randomly cropped from the augmented data.
With respect to the submission of batches and weight initialisation, models were trained identically.

\section{Results and discussion}

\begin{figure}
\includegraphics[width=\textwidth]{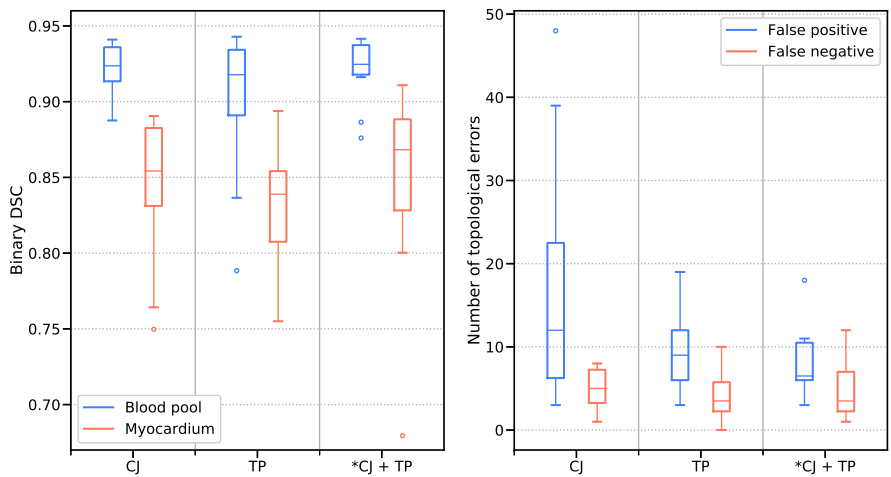}
\caption{CNN segmentation performance in terms of spatial overlap (left) and topological accuracy (right), where training data are augmented by spatial transformation which is: CJ - clinically justified; TP - topology-preserving; CJ + TP - clinically justified and topology-preserving *using our novel augmentation pipeline.} \label{boxplots}
\end{figure}

Our results assess the impact of two characteristics of data augmentation: (i) whether spatial transformation is clinically informed; and (ii) whether label map topology is preserved after transformation and resampling.

Perhaps for their ease of implementation, previous work has employed orthogonal rotations \cite{wolterink17,yu17,yu17-frAQB} and lateral inversion \cite{yu17,yu17-frAQB} only.
Whilst these transformations are guaranteed to preserve label map topology, they are not representative of the distribution of CMR data seen in the clinic.
For example, since patient position is invariably head first-supine, orthogonal rotations are unrealistic. 
Though small variations in patient orientation are observed, these are best captured by small angle rotations. We seek to describe this distribution through the use of the clinically justified transformations shown in figure \ref{pipeline}.

With respect to spatial overlap, figure \ref{boxplots} suggests a benefit to this approach. 
The DSC improves from $0.918\ (0.891, 0.934)$\footnote{All results reported as median (interquartile range).} to $0.925\ (0.918, 0.938)$ and from $0.839\ (0.808, 0.854)$ to $0.868\ (0.828, 0.888)$, for the blood pool and myocardium classes respectively.
These results suggest that unlike orthogonal rotation, clinically justified transformations act not only as a source of regularisation but are also beneficial to the generalisability of the network.
Overall, these results are consistent with, or in the case of the myocardium class, exceed those achieved by previous work \cite{li17,wolterink17,yu17,yu17-frAQB}.

As well as spatial overlap, we are also concerned with the topology of the inferred blood pool label map.
Of particular interest are the false positive clusters which characterise defective interface segmentations and the anomalous presence of defects.
Figure \ref{boxplots} demonstrates that compared to previous work, the naive introduction of clinically justified transformations has a detrimental effect on inferred blood pool topology.
The median number of topologically relevant false positive clusters increases from $9.0\ (6.0, 12.0)$ to $12.0\ (6.3, 22.5)$.
However, when coupled with our topology-preserving augmentation pipeline and its consideration of CCT, this number falls dramatically to $6.5\ (6.0, 10.5)$: a statistically significant improvement according to Wilcoxon Signed Rank test ($p=0.022$).
This also represents improved performance compared with the clinically unrealistic though topology-preserving augmentation schemes used in previous work.

These observations suggest that the topological features of inferred cardiac segmentations are strongly dependent on the  training data.
Though perhaps predictable, it must be remembered that in the context of complex morphology, topological features can be encapsulated by relatively few voxels.
In fact, across the ten thousand augmented representations of the ten ground truth label sets we produced, less than 0.5\% of blood pool voxels were changed by our topology-preserving pipeline.

Figure \ref{boxplots} shows that best performance can be attributed to augmentation pipelines which are both clinically justified and which preserve ground truth label topology.
Our topology-preserving augmentation pipeline provides a means of simultaneously achieving both qualities.

\section{Conclusion}

Our work demonstrates for the first time the importance of label map topology to the task of CNN-based CHD segmentation from 3D CMR images of paediatric patients.
We have presented a novel pipeline for the augmentation of training data for CNN optimisation.
Invoking the concept of CCT and developing an associated, clinically meaningful metric, we show that the properties of this pipeline - allowing for clinically justified data augmentation whilst preserving arbitrary label map topology - are beneficial to the topological properties of inferred segmentations. 
We speculate that these findings may be applicable to any medical image segmentation task for which morphologically complex foreground objects can be represented as a number of contiguous sub-classes.

%
%
%
\bibliographystyle{splncs04}
\bibliography{refs}

\begin{thebibliography}{1}
\providecommand{\url}[1]{\texttt{#1}}
\providecommand{\urlprefix}{URL }
\providecommand{\doi}[1]{https://doi.org/#1}

\bibitem{bazin07}
Bazin, P.L., Pham, D.L.: Topology correction of segmented medical images using
  a fast marching algorithm. Computer methods and programs in biomedicine
  \textbf{88}(2),  182--190 (2007)

\bibitem{li17}
Li, J., Zhang, R., Shi, L., Wang, D.: Automatic Whole-Heart Segmentation in
  Congenital Heart Disease Using Deeply-Supervised 3D FCN, pp. 111--118.
  Springer International Publishing, Cham (2017).
  \doi{10.1007/978-3-319-52280-7\_11}

\bibitem{losel17}
Losel, P., Heuveline, V.: A GPU Based Diffusion Method for Whole-Heart and
  Great Vessel Segmentation, pp. 121--128. Springer International Publishing,
  Cham (2017). \doi{10.1007/978-3-319-52280-7\_12}

\bibitem{milletari16}
Milletari, F., Navab, N., Ahmadi, S.A.: V-net: Fully convolutional neural
  networks for volumetric medical image segmentation. In: 3D Vision (3DV), 2016
  Fourth International Conference on. pp. 565--571. IEEE

\bibitem{pace15}
Pace, D.F., Dalca, A.V., Geva, T., Powell, A.J., Moghari, M.H., Golland, P.:
  Interactive whole-heart segmentation in congenital heart disease. Med Image
  Comput Comput Assist Interv  \textbf{9351},  80--88 (2015).
  \doi{10.1007/978-3-319-24574-4\_10}

\bibitem{wolterink17}
Wolterink, J.M., Leiner, T., Viergever, M.A., Išgum, I.: Dilated Convolutional
  Neural Networks for Cardiovascular MR Segmentation in Congenital Heart
  Disease, pp. 95--102. Springer International Publishing, Cham (2017).
  \doi{10.1007/978-3-319-52280-7\_9}

\bibitem{yu17}
Yu, L., Cheng, J.Z., Dou, Q., Yang, X., Chen, H., Qin, J., Heng, P.A.:
  Automatic 3d cardiovascular mr segmentation with densely-connected volumetric
  convnets. In: International Conference on Medical Image Computing and
  Computer-Assisted Intervention. pp. 287--295. Springer (2017)

\bibitem{yu17-frAQB}
Yu, L., Yang, X., Qin, J., Heng, P.A.: 3D FractalNet: Dense Volumetric
  Segmentation for Cardiovascular MRI Volumes, pp. 103--110. Springer
  International Publishing, Cham (2017). \doi{10.1007/978-3-319-52280-7\_10}

\bibitem{zuluaga17}
Zuluaga, M.A., Biffi, B., Taylor, A.M., Schievano, S., Vercauteren, T.,
  Ourselin, S.: Strengths and Pitfalls of Whole-Heart Atlas-Based Segmentation
  in Congenital Heart Disease Patients, pp. 139--146. Springer International
  Publishing, Cham (2017). \doi{10.1007/978-3-319-52280-7\_14}

\end{thebibliography}

\end{document}